\documentclass[letter,longauth,traditabstract]{aa} 
\usepackage{graphicx}
\usepackage{txfonts}
\begin{document}
   \title{$Herschel$ discovery  of a new  class of cold,  faint debris
     discs\thanks{Herschel  is an ESA  space observatory  with science
       instruments  provided  by  European-led Principal  Investigator
       consortia and with important participation from NASA.}}


    \author{C. Eiroa
            \inst{1} 
            \and  J. P. Marshall\inst{1}
    \and A. Mora\inst{2}          
\and A. V. Krivov\inst{3}
\and B. Montesinos\inst{4}
\and O. Absil\inst{5}
\and D. Ardila\inst{6}
\and M. Ar\'evalo\inst{4}
\and J.-Ch. Augereau\inst{7}
\and A. Bayo\inst{8}
\and W. Danchi\inst{9}
\and C. del Burgo\inst{10}
\and S. Ertel\inst{11}
\and M. Fridlund\inst{12}
\and B.M. Gonz\'alez-Garc\'\i a\inst{13}
\and A. M. Heras\inst{12}
\and J. Lebreton\inst{7}
\and R. Liseau\inst{14}
\and J. Maldonado\inst{1}
\and G. Meeus\inst{1}
\and  D. Montes\inst{15}
\and G.L. Pilbratt\inst{12}
\and A. Roberge\inst{9}
\and J. Sanz-Forcada\inst{4}
\and K. Stapelfeldt\inst{16}
\and P. Th\'ebault\inst{17}
\and G. J. White\inst{18,19}
\and S. Wolf\inst{11}
 %
           }
   \institute{Dpt. F\'\i sica Te\'orica, Facultad de Ciencias, 
            Universidad Aut\'onoma de Madrid,
            Cantoblanco, 28049 Madrid, Spain
             \email{carlos.eiroa@uam.es}
             \and ESA-ESAC Gaia SOC. P.O. Box 78, E28691 Villanueva de la Ca{\~n}ada, Madrid, Spain
\and   Astrophysikalisches Institut und Universit{\"a}tssternwarte,   
 Friedrich-Schiller-Universit{\"a}t,
 Schillerg{\"a}{\ss}chen 2-3, 07745 Jena, Germany
\and Dpt. de Astrof\'\i sica, Centro de Astrobiolog\'\i a (INTA-CSIC), 
ESAC Campus, P.O.Box 78, E-28691 Villanueva 
de la Ca{\~n}ada, Madrid, Spain
\and Institut d'Astrophysique et de G\'eophysique, Universit\'e de Liege, 17 
All\'ee du Six Ao\^ut, B-4000 Sart Tilman, Belgium
\and NASA Herschel Science Center, California Institute of Technology, 1200 E. 
California Blvd., Pasadena, CA 91125, USA
\and Universit\'e Joseph Fourier/CNRS, Laboratoire d'Astrophysique de Grenoble,
 UMR 5571, Grenoble, France
\and European Space Observatory, Alonso de Cordova 3107, Vitacura, Casilla 
19001, Santiago 19, Chile
\and NASA Goddard Space Flight Center, Exoplanets and Stellar Astrophysics, 
Code 667, Greenbelt, MD 20771.USA 
\and UNINOVA-CA3, Campus da Caparica, Quinta da Torre, Monte de Caparica,
2825-149 Caparica, Portugal 
\and Christian-Albrechts-Universit\"at zu Kiel, Institut f\"ur Theoretische 
Physik und Astrophysik, Leibnizstr. 15, 24098 Kiel, Germany 
\and ESA Astrophysics \& Fundamental Physics Missions Division, ESTEC/SRE-SA,  
Keplerlaan 1, NL-2201 AZ Noordwijk,
The Netherlands
 \and INSA at ESAC, E-28691 Villanueva de la Ca{\~n}ada, Madrid, Spain
\and Onsala Space Observatory, Chalmers University of Technology, Se-439 92 Onsala, Sweden
 \and Universidad Complutense de Madrid, Facultad de Ciencias F\'\i sicas, Dpt.
 Astrof\'\i sica, Av. Complutense s/n, 28040 Madrid, Spain
\and Jet Propulsion Laboratory, California Institute of Technology, MS 183-900, 4800 Oak Grove Drive, 
Pasadena, CA 91109, USA
\and LESIA, Observatoire de Paris, 92195 Meudon France 
\and  Department of Physics and Astrophysics, Open University, Walton Hall, 
Milton Keynes MK7 6AA, UK
\and Rutherford Appleton Laboratory, Chilton OX11 0QX, UK
}

\abstract
{We  present $Herschel$ PACS  100 and
160 $\mu$m observations of the solar-type stars $\alpha$ Men, HD 88230
and HD 210277, which form part of the FGK stars sample of the Herschel
Open  Time  Key  Programme  (OTKP)  $DUNES$  ($DU$st  around  $NE$arby
$S$tars).  Our observations show small infrared excesses at 160 $\mu$m
for  all three  stars. HD  210277  also shows  a small  excess at  100
$\mu$m, while the 100 $\mu$m  fluxes of $\alpha$ Men and HD 88230 agree
with  the  stellar   photospheric  predictions.   We  attribute  these
infrared excesses to a new class of cold, faint debris discs. $\alpha$
Men and HD 88230 are spatially resolved in the PACS 160 $\mu$m images,
while  HD 210277  is  point-like at  that  wavelength.  The  projected
linear sizes of the extended emission lie in the range from $\sim 115$
to $\leq$ 250 AU.  The  estimated black body temperatures from the 100
and  160 $\mu$m  fluxes  are  $\lesssim$ 22  K,  while the  fractional
luminosity of the cold  dust is $L_{dust}/L_\star \sim 10^{-6}$, close
to  the luminosity of  the Solar-System's  Kuiper belt.   These debris
discs  are the  coldest and  faintest discs  discovered so  far around
mature stars  and cannot easily  be explained by  invoking "classical"
debris disc models.}

  \maketitle
%

\section{Introduction}

Debris  discs are  tenuous  structures associated with main sequence  
stars formed by  second
generation  dust, which  has  resulted from  the  collisions of  solid
bodies continuously supplying the circumstellar environment with small
dust particles.  This formation sequence is inferred from the lifetime
of the dust  grains against destructive collisions, Poynting-Robertson
drag and  radiation pressure, which is  much shorter than  the ages of
the host stars. These discs  are visible in reflected light at optical
wavelengths and in thermal  radiation at mid-/far-IR and submillimeter
wavelengths (Aumann et al.,  1984; Backman \& Paresce, 1993).  
General  disc characteristics  are  grain black  body temperatures  of
$\sim 50  - 100$ K,  fractional     luminosities      $f >
L_{dust}/L_{\star}\sim10^{-5}$, and radii  from less than 10 AU to several
times 100 (e.g.    Absil et al. 2006, Su et al. 2005, Trilling   et
al. 2008).  Debris discs are  considered analogues of the Solar System
asteroid  and Kuiper  belts, although  their luminosities  are usually
more than  100 times the  Kuiper belt level of  $L_{dust}/L_{Sun} \sim
10^{-7}  - 10^{-6}$  (Stern 1996,  Vitense et  al. 2010).   The
sensitivity  of  the 3.5  m  $Herschel$  far-infrared space  telescope
(Pilbratt  et  al.   2010)  with  its instrument  PACS  (Poglitsch  et
al. 2010)  offers the possibility of characterising  colder ($\sim 30$
K)  and fainter  ($L_{dust}/L_{\star}  $ few  times $10^{-7}$)  debris
discs  with spatial  resolution  $\sim$  60  AU (FWHM) at 10  pc, 
i.e.,  true extra-solar Kuiper belts.


DUNES  is  a  $Herschel$  OTKP  designed to  detect  and  characterise
extra-solar  analogues to  the Kuiper  belt around  main  sequence FGK
nearby  stars (Eiroa  et al.  2010).  In  this letter  we  present the
results for three stars from the DUNES sample: $\alpha$ Men, HD 88230,
and  HD  210277  as  clear  examples  of  the  advantages  offered  by
$Herschel$ observations: they trace a  new class of cold, $T \lesssim$
22  K,  spatially  resolved  debris  discs with  very  low  fractional
luminosities.  These discs have remained unobserved by previous far-IR
and submillimeter studies.   Table  1 gives some properties of the
  stars.  Ages  are based on  the $\log$ R$'_{HK}$ activity  index and
  have an uncertainty of 60\% (Mamajek \& Hillenbrand 2008). HD 210277
  hosts  a Jupiter-like  planet (Marcy  et al.   1999). This  star and
  $\alpha$  Men have  faint stellar  companions, but  neither  the
  measurements nor the photospheric  predictions are affected by them.
  Eiroa et  al.  {\it in prep}  will present a full  discussion of the
  stars and the general results of the DUNES survey.

%
%

\section{Observations and data reduction}

\begin{table}[t]
\label{stars}
\caption{Stellar properties.}
\begin{center}
\begin{tabular}{llll}
\hline

Star                &   $\alpha$ Men &   HD 88230 &  HD 210277 \\
\hline
Sp. Type            &   G5V-G7V       &  K6V-M0V       &  G0V, G7V-G9V \\
Distance (pc)       & 10.2      & 4.9       & 21.6 \\ 
$L_* (L_\odot)$      & 0.85      & 0.15      & 1.10\\
T$_{eff}$(K)         & 5590      & 3850      & 5540 \\
Age (Gyrs)  & 5.5       & 6.6         & 6.9      \\
\hline            
\end{tabular}
\end{center}
\end{table}


\begin{table}
\label{AORs}
\caption{Log of the PACS 100 $\mu$m and 160 $\mu$m observations.}
\begin{tabular}{lllrr}
\hline
Star    & HIP   & Obs. ID   & OT (sec) \\ 
\hline
$\alpha$ Men  &  29271 &1342203297/8  & 1116 \\
$\alpha$ Men  &        & 1342216043/4 & 2244  \\
HD 88230  &  49908 & 1342210610/1 & 4500   \\
HD 210277 & 109378 & 1342211126/7 & 4500 \\
\hline
\end{tabular}
\end{table}

$\alpha$ Men,   HD~88230  and  HD~210277   were  observed   with  PACS
100/160~$\mu$m in scan map mode.  For each star, two scans at position
angles  70$^{\circ}$ and  110$^{\circ}$  were carried  out, each  scan
consisting of  10 legs  with separation of  4$''$, length of  3$'$ and
medium  speed  of 20$''$/s.  Table  2  gives  the scan  identification
numbers  (Obs.  ID)  and   the  total  duration  of  the  observations
(OT). $\alpha$ Men was  observed twice in order to  increase the S/N
ratio.  Data  reduction  was   made  using  the  Herschel  Interactive
Processing Environment  (HIPE) version 7.2. The  individual scans were
mosaiced to produce  the final image at each  band. To check the
consistency of the reduction  and analysis (particularly regarding the
effect of correlated noise), mosaics  were produced at both the native
3$\farcs$2 for 100~$\mu$m (green) and  6$\farcs$4 for 160~$\mu$m (red), as 
well as super-sampled pixel scales 1.0$''$ (green) and 2.0$''$ (red), the
latter being the  default pixel size in HIPE for this  type of data.  The 
images with the native pixel size scales avoid, at least partly, the 
correlated noise in the PACS images\footnote{Technical Note PICC-ME-TN-037 in http://herschel.esac.esa.int} (see also Fruchter \& Hook, 2002).  A
high-pass filter  was used to  remove large scale  background emission
from the images, with filter widths  of 15$''$ and 25$''$ in the green
and  red  channels. To  prevent the removal  of  any extended structure 
near the stars, regions where the sky
brightness  exceeded a  threshold value  in the  image defined  by the
standard  deviation of  all the  positive pixels  ($S~>~10^{-6} Jy/pixel$) were
masked from that process.  Absolute flux calibration uncertainties are
$\sim$3\% and $\sim$5\% for the green and red bands (see technical note below).

\section{Results}

Table ~\ref{pointing} gives the J2000.0 optical equatorial coordinates
of the stars as well as  their 100 $\mu$m peak positions corrected for
the proper motions  of the stars (van Leeuwen,  2007). Offsets between
the optical and the PACS positions (column 4 of Table ~\ref{pointing})
are  within  $\sim$  1.5$\sigma$  the Herschel  pointing  accuracy  of
2$\farcs$4   in  this   observing   mode  (S\'anchez-Portal,   private
communication).

Fig. 1 shows the 100 and 160 $\mu$m images and isocontour plots of the
stars.  $\alpha$ Men  and HD 88230 are point-like  at 100 $\mu$m, FWHM
$\sim$ 6$\farcs$3 $\times$ 6  $\farcs$5, while both stars are resolved
at   160   $\mu$m,   with   angular   sizes/position   angles   $\sim$
18$\arcsec$/52$\degr$      and      $\sim$      23$\farcs$5/45$\degr$,
respectively. The  extended emission appears clearly  asymmetric in HD
88230, with the  star located at the North-Eastern  side.  None of the
objects  are resolved  in  the direction  orthogonal  to the  extended
emission.  HD 210277 is unresolved at both wavelengths.


PACS fluxes (Table  ~\ref{pointing}) have been estimated  using circular 
and rectangular  aperture photometry, taking  special care  to  choose
the reference  background region due to the presence of  field objects. 
Specifically, measurements of HD 210277 take into account the presence of 
the bright object located  at $\sim$16$\arcsec$ North-East from the  star; 
in the case of  this star we have,  in addition, carried  out PSF photometry
using  the DAOPHOT software  package.  The  PSF photometry  fluxes are
$F$(100) = 8.4$\pm$0.3 mJy and $F$(160) = 14.3$\pm$0.4 mJy (errors are
those  of  the PSF  fits),  which  are  consistent with  the  aperture
photometry estimates. 
Errors have been estimated using  a variety of methods, including circular and 
rectangular boxes at different positions in the nearby fields. Sky sizes for 
the errors estimates are equal to the area of the aperture used for the 
photometry for all three stars and both bands; in particular the sizes of 
the extended emission around $\alpha$ Men and HD 88230.

PACS  fluxes  have been  compared  to  predicted stellar  photospheric
fluxes  (Table ~\ref{pointing})  using Gaia/PHOENIX  models  (Brott \&
Hauschildt, 2005),  with the stellar  parameters as given in  Eiroa et
al.  (in prep). Fig. 1  shows the spectral energy distributions (SEDs)
of the  stars, where  PACS fluxes are  plotted together  with optical,
near-IR,  IRAS,  AKARI,  and  {\em  Spitzer}/MIPS  and  IRS  data;  in
addition, the best $\chi$-square photospheric fit is shown.  To assess
the presence of an excess at  100 and/or at 160 $\mu$m we require that
the observed fluxes, $F_{\rm PACS}$,  exceed by at least 3$\sigma$ the
predicted photospheric fluxes, $F_*$  ($\chi_\lambda = (F_{\rm PACS} -
F_*)/\sigma_\lambda$).
No excesses are detected at 100  $\mu$m for $\alpha$ Men and HD 88230,
while it  is seen  in HD 210277.  All three  stars do show  160 $\mu$m
excesses. We also note that the  SED slopes from 100 to 160 $\mu$m are
$\alpha$ = - 0.5$\pm$0.7 ($\alpha$  Men), $\alpha$ = - 0.7$\pm$0.6 (HD
88230), and  $\alpha$ = 0.8$\pm$0.8 (HD 210277),  which clearly differ
from the  expected Rayleigh-Jeans  behaviour ($\alpha$ =  - 2.0)  of a
stellar photosphere in this wavelength regime.

\begin{table*}[ht]
\label{pointing}
\caption{Optical positions of the stars and of their PACS 100 identified 
counterparts. Observed PACS fluxes with 1$\sigma$ statistical errors 
($F_{\rm PACS}$), and predicted photospheric fluxes ($F_{\star}$). 
Flux units are mJy.}
\begin{center}
\begin{tabular}{lcccrrrrrr}
\hline
Star          & Optical position & PACS 100 $\mu$m position & Offset
& \multicolumn{3}{r}{PACS 100 $\mu$m}
& \multicolumn{3}{r}{PACS 160 $\mu$m} \\
        & (J2000.0)      & (J2000.0) & $\arcsec$
& $F_{\rm PACS}$ & $F_{\star}$ & $\chi_{100}$
& $F_{\rm PACS}$ & $F_{\star}$ & $\chi_{160}$\\
\hline 
$\alpha$ Men & 06 10 14.47  -74 45 11.0 & 06 10 14.53 -74 45 11.1 & 0.3 &17.8$\pm$1.3 & 18.0 &-0.2 & 14.4$\pm$2.0 &  7.0 &3.7 \\
  HD 88230   & 10 11 22.14  +49 27 15.3 & 10 11 21.88 +49 27 17.2 & 3.2 &22.5$\pm$0.9 & 25.7 &-3.6 & 16.0$\pm$1.7 & 10.0 &3.5 \\  
  HD 210277  & 22 09 29.87  -07 32 55.2 & 22 09 30.08 -07 32 54.1 & 3.3 & 8.5$\pm$1.0 &  4.6 & 3.9 & 12.4$\pm$1.6 &  1.8 &6.6 \\  
\hline
\end{tabular}
\end{center}
\end{table*}

\begin{figure*}
\label{contornos}
\centering
\includegraphics[width=4.7cm]{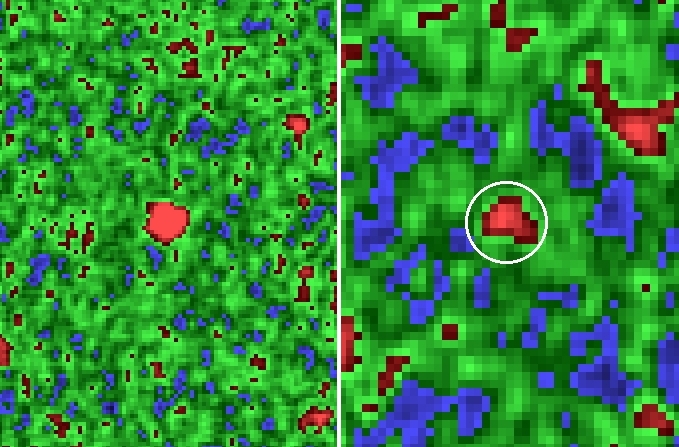}
\hspace{0.7cm}\includegraphics[width=4.7cm]{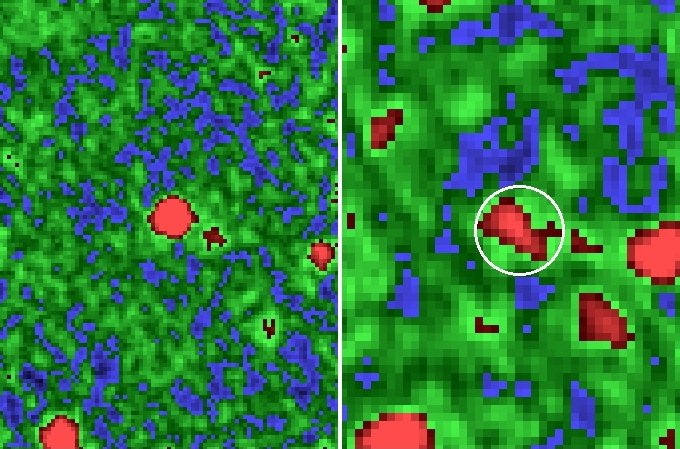}
\hspace{0.7cm}\includegraphics[width=4.7cm]{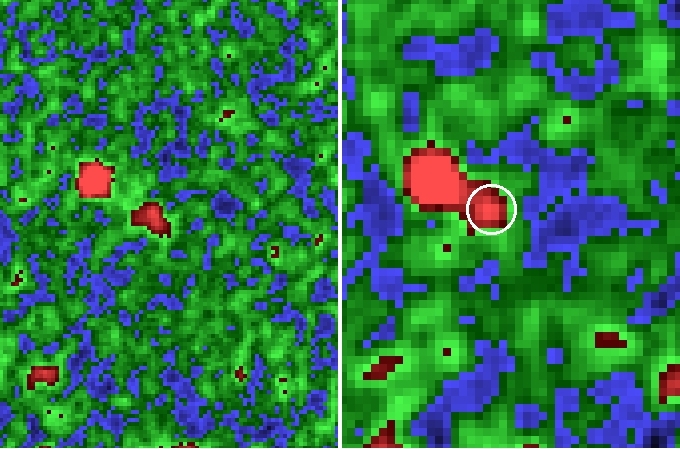}\\
\includegraphics[width=4.7cm,angle=-90]{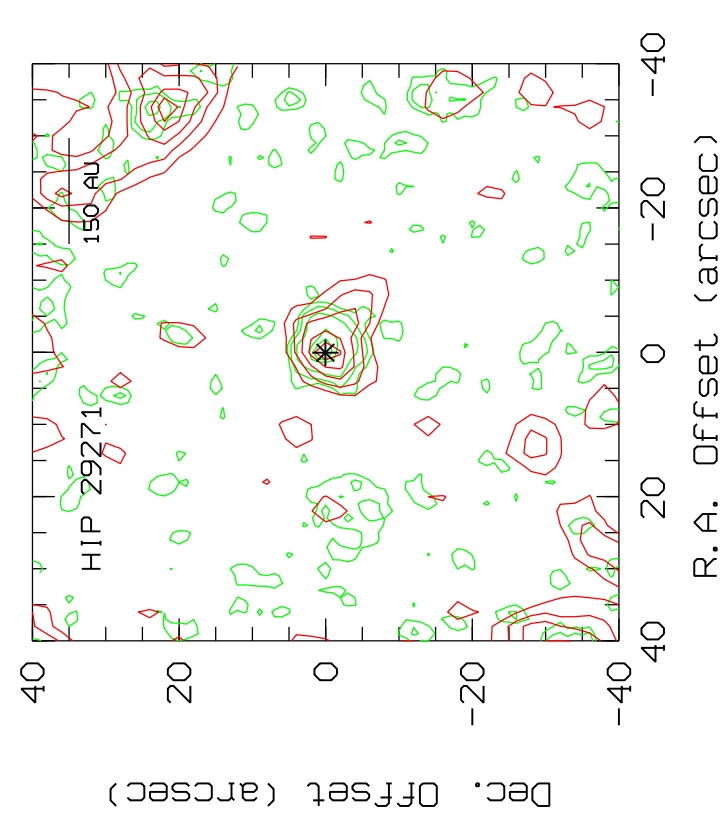}
\includegraphics[width=4.7cm,angle=-90]{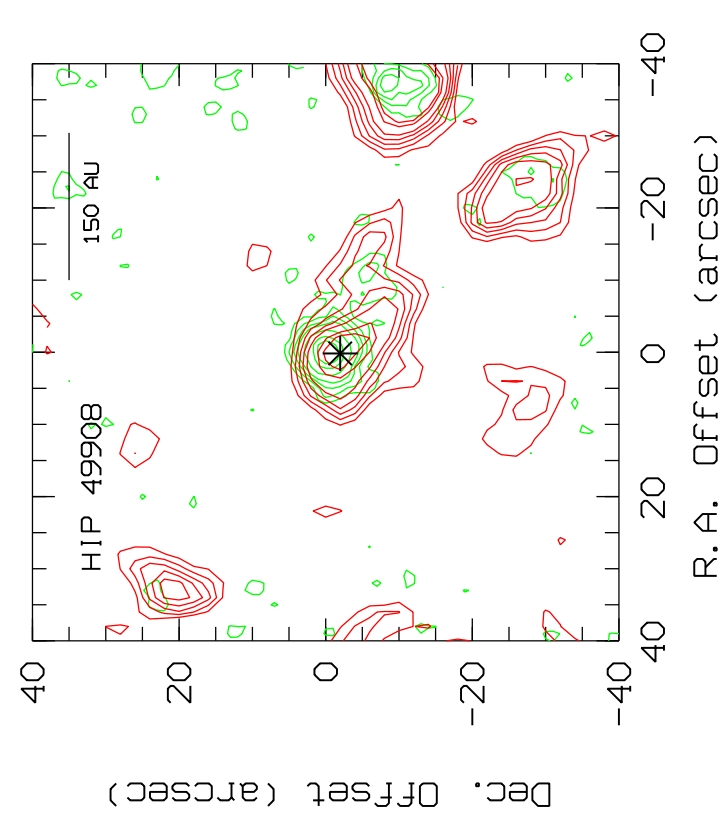}
\includegraphics[width=4.7cm,angle=-90]{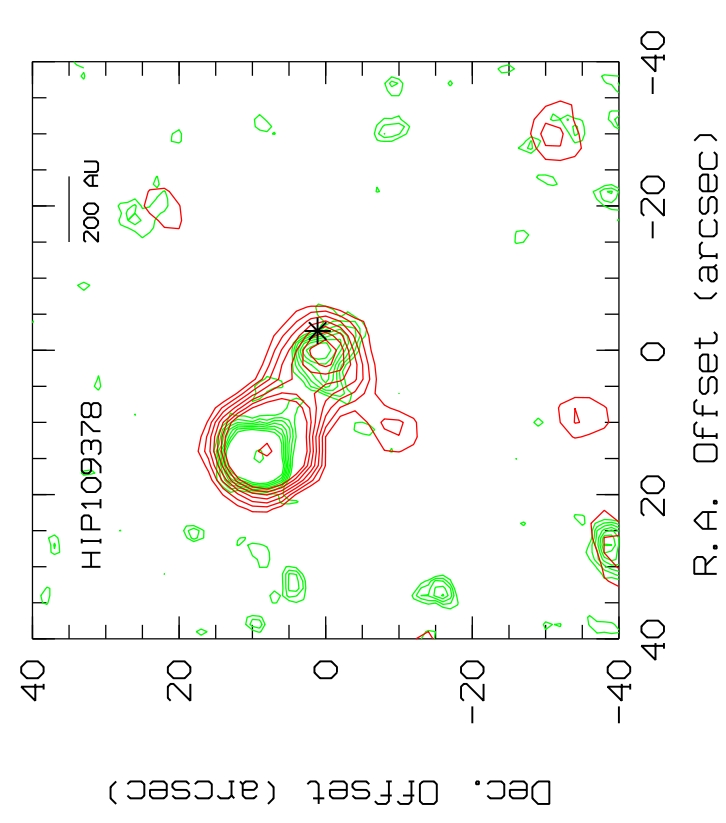}\\
\includegraphics[width=5cm]{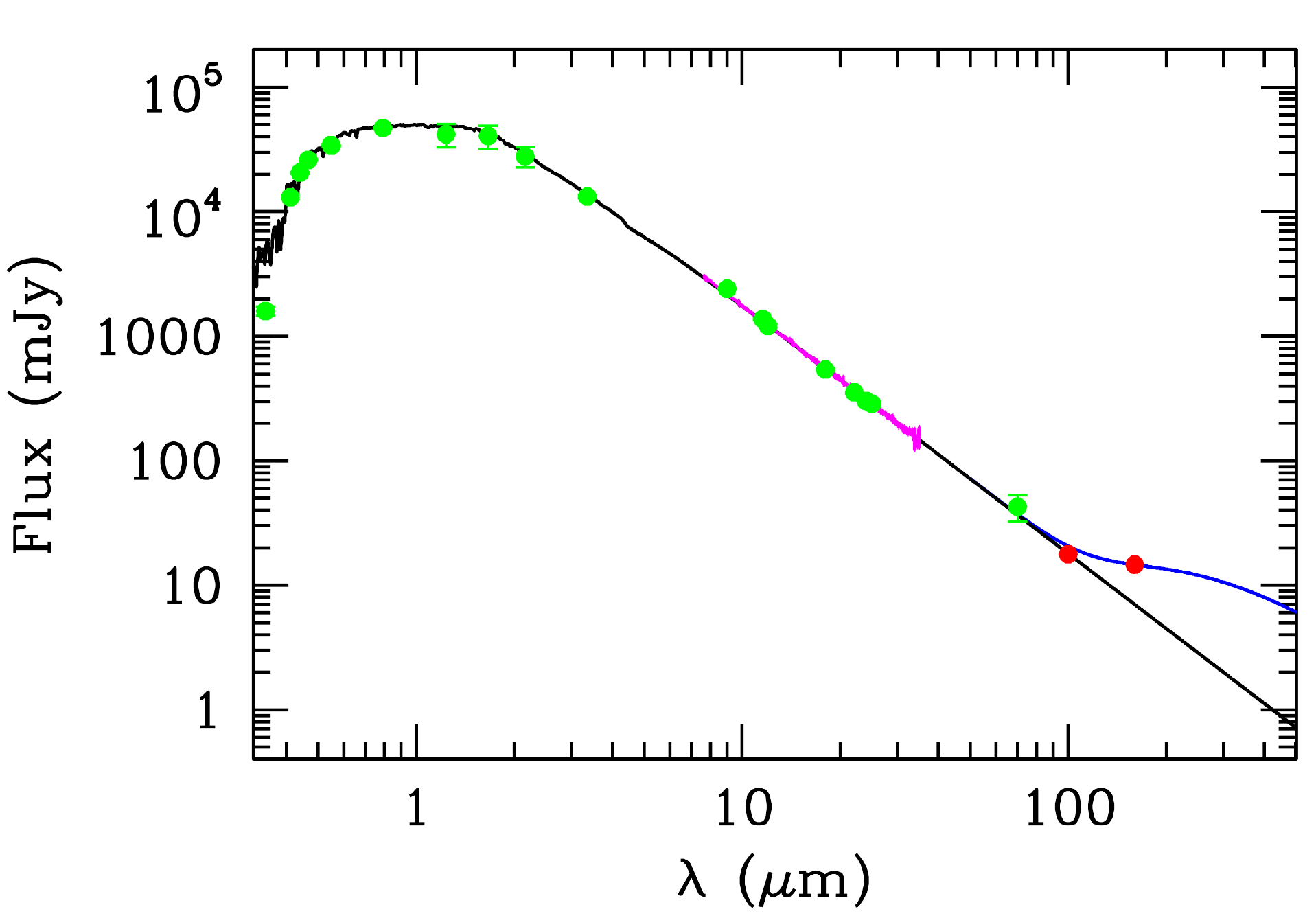}
\hspace{0.7cm}\includegraphics[width=5cm]{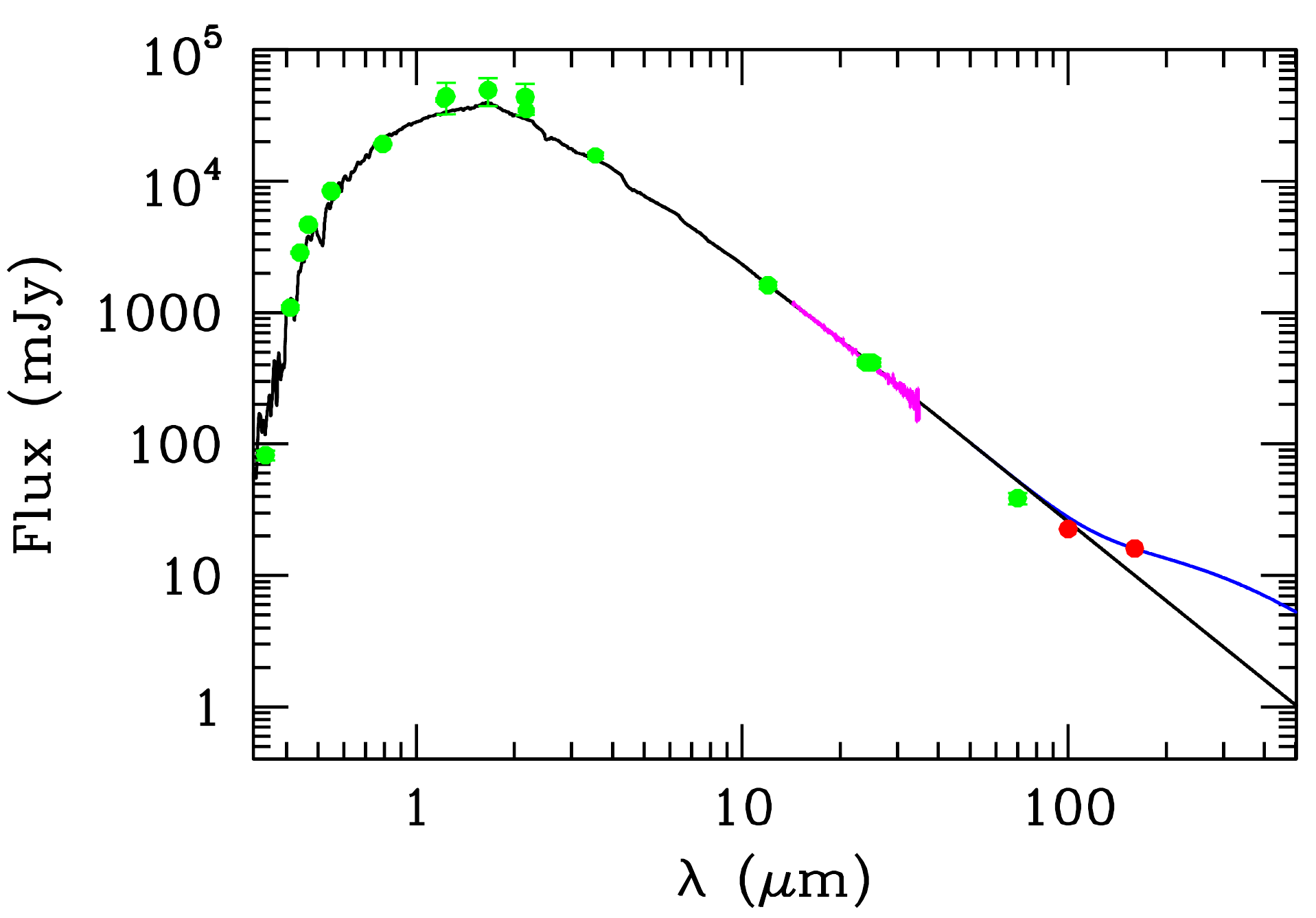}
\hspace{0.7cm}\includegraphics[width=5cm]{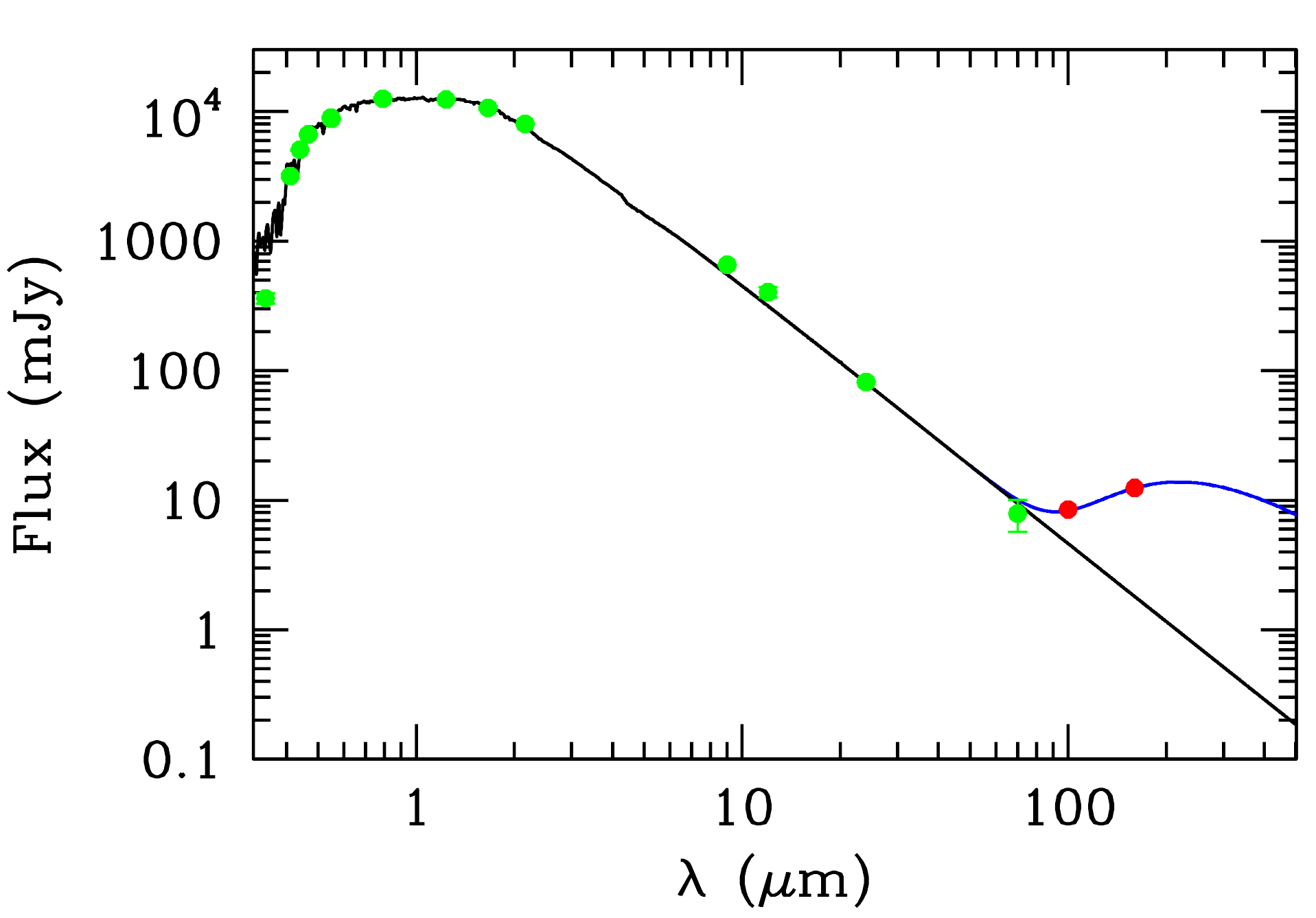}
\caption{PACS images, isocontours  and SEDs of
  $\alpha$ Men/HIP  29271 (left), HD  88230/HIP 49908 (middle)  and HD
  210277/HIP 109378  (right).  Images:  100 $\mu$m (left),  160 $\mu$m
  (right). North is  up and East to the  left. Isocontours: 100 $\mu$m
  contours are in  green colour while the 160 $\mu$m  ones are in red.
  $\alpha$ Men: 100  $\mu$m contours (10, 20, 40, 80 and  90 \% of the
  flux  peak); 160  $\mu$m contours  (20,40,60,80, 90  \% of  the flux
  peak) HD 882308: 100 $\mu$m contours ( 10, 20,30,40,60, 80, 90 \% of
  the flux peak);  160 $\mu$m contours (20, 30, 40, 50,  60, 80, 90 \%
  of the flux peak) HD 210277:  100 $\mu$m contours ( (30, 40, 50, 60,
  70, 80,  90 \% of the flux  peak); 160 $\mu$m contours  (50, 60, 70,
  80, 90  \% of  the flux peak).  The lowest  contour in all  cases is
  $\approx 3 \sigma$.  The optical position of the stars are indicated
  by the  symbol ``*'' in the  isocontour plots; a segment
  indicates the projected linear sizes at the distance of each star. SED plots: black line is the
  photospheric fit while the blue line is the photosphere plus a 22 K black body.}
\end{figure*}

\subsection{Comments on the PACS images}

All three fields show $\sim$  2-3 red sources per square arcmin, which
are likely to be background  galaxies. In particular, there is a faint
peak at  $\sim$13$\arcsec$ towards the SW  from HD 88230  and a bright
one  at $\sim$16$\arcsec$  towards the  NE  from HD  210277.  We  have
consulted   the  NASA/IPAC  extragalactic   database  to   search  for
counterparts  without  finding any  association.   Although we  cannot
firmly  exclude  a  coincidental   alignment  or  contamination  of  a
background source(s) in the line of  sight of our stars, we think that
it is  unlikely for the $Herschel$  sources presented here  due to the
close correlation  between the  optical and $Herschel$  positions, and
the photospheric  predictions and the estimated  $Herschel$ 100 $\mu$m
fluxes.  In fact,  following in a first approach  the  source counts  by Berta  et
al. (2010),  the average density of extragalactic  sources with fluxes
$\sim$ 6-7  mJy and $\sim$ 12  mJy - i.e., the  measured excesses at 160 
$\mu$m from
$\alpha$  Men/HD 88230  and HD  210277, respectively  (Table 3)  - are
2/arcmin$^2$ and 0.7/arcmin$^2$. Thus, given the optical/160 $\mu$m offsets 
(Fig. 1) the a priori probability of an
accidental alignment is very small, clearly smaller than 5\%. We refer
to a  future paper (del Burgo et  al., {\it in prep.})  for a detailed
study on source contamination within the DUNES fields.

\section{Analysis}

The small PACS excesses  above the photospheric fluxes are interpreted
as  due  to cold  debris  discs around  the  stars.   Black body  dust
temperatures, $T_{\rm  dust}$, can be  estimated from the 100  and 160
$\mu$m excess fluxes for HD 210277; in the case of $\alpha$ Men and HD
88230, an upper limit for  $T_{\rm dust}$ can be calculated taking the
100 $\mu$m  flux as 3$\sigma$  statistical noise.  $T_{\rm  dust}$ for
the three stars is $\lesssim$  22 K (Table 4); the corresponding inner
radii  of  discs  with  black  body grains  at  this  $T_{\rm  dust}$,
considering  the luminosity  of the  stars,  are also  given in  Table
4. Fig.  1 shows the excellent  agreement between the  combined SED of
the stellar photospheric fits plus  22 K black bodies and the observed
SEDs.

Deconvolution of the  images can be used to estimate  the true size of
the resolved  discs in $\alpha$ Men and HD 88230. 
Our method  first removes  the photospheric
contributions  from each image  by subtracting  a  PSF with  a peak
scaled to  the predicted  photospheric flux level.  The PSF model 
uses $\alpha$ Bo\"otis
images, rotated
to match  the roll angle of  the telescope during  the observations of
the   DUNES  stars.   After  star subtraction,  the  images  are
deconvolved using both modified Wiener and Richardson-Lucy algorithms.  
The noise model takes
into  account that the main contributor  is the  telescope thermal
emission.  Both algorithms  produce consistent  results  although with
different noise patterns.  Fig. 2 shows the 160 $\mu$m star-subtracted
and  the  Wiener-deconvolved images  of  $\alpha$  Men  and HD  88230;
estimated  sizes  of  the  deconvolved  sources  are  16$\arcsec$  and
21$\arcsec$,  respectively.  Table  4 gives  the linear  sizes  of the
semi-major axes ($\sim$3$\sigma$ contours)  from both the original and
deconvolved  PACS 160 $\mu$m  images. In the case
of HD 210277, 
the value in Table  4   corresponds  to  an  upper  limit   of  12$\arcsec$,  i.e.,
approximately the  160 $\mu$m beam size. The comparison of both the
directly  observed and  deconvolved semi-major  axes to  the estimated
radii from  $T_{\rm dust}$ clearly  indicates that the  observed discs
(assuming as  a first approach  symmetric discs) are smaller  than the
expected sizes for black body discs.

\begin{figure}
\label{deconvolution}
\centering
\includegraphics[width=6.75cm]{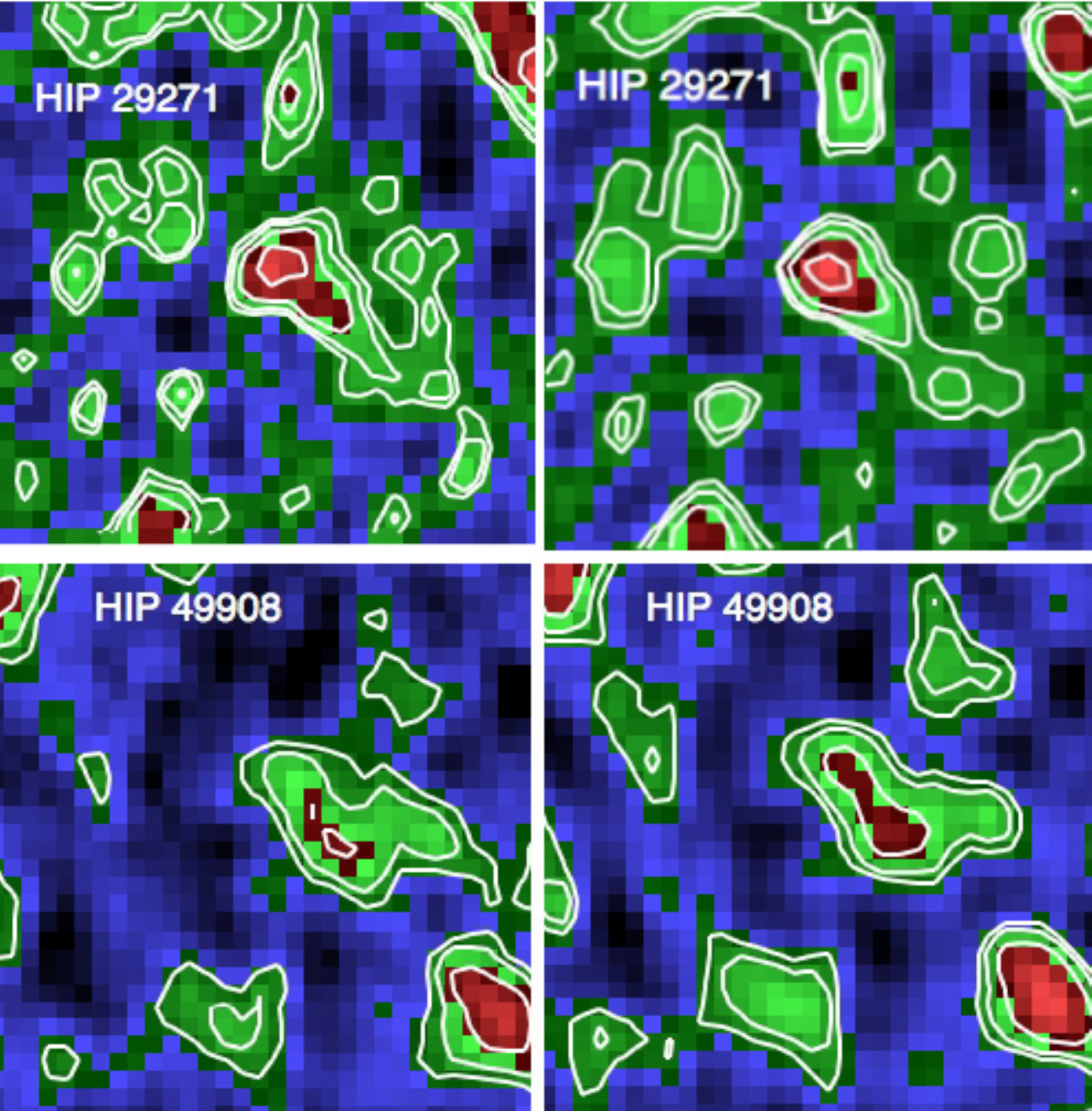}

\caption{160 $\mu$m star-subtracted (left) and deconvolved (right) images. 
     Up: $\alpha$ Men/HIP 29271. Contours: 10\%, 20\%,40\%, 80\% of the peak.  
     Bottom:  HD 88230/HIP 49908. Contours: 20\%,40\%, 80\% of the peak. 
     North is up and East to the left.  Field size is $60\arcsec \times 60 \arcsec$.}
\end{figure}

Dust fractional  luminosities,  $f$  (Table 4), can be obtained by taking  the  160 $\mu$m fluxes
and assuming  the star  temperatures given  in Table 1  and $T_{\rm dust}$  
= 22  K (Beichman et al., 2006).   The estimated  $f$  values are 
of  the order  of 10$^{-6}$, close  to the
Kuiper belt level.  Essentially similar  $f$ values are obtained 
taking the maximum  wavelength corresponding to 22 K,  i.e., 230 $\mu$m,
and  its expected  flux extrapolating  from  the one  measured at  160
$\mu$m.

\section{Discussion}

\begin{table}[h!]
\label{dust}
\caption{Black body dust temperatures from the far-IR excesses, $T_{\rm dust}$ and the corresponding 
estimated radii.  Observed 3$\sigma$ linear sizes 
  (semi-major axes) of the extended emission at 160 $\mu$m 
in the original and deconvolved images. $f$ is the dust fractional luminosity.}
\begin{center}
\begin{tabular}{lrrrrc}
\hline
Star        & $T_{\rm dust}$   &  Size       & Size                & Size                    & $f$\\
            &                &  Est.  & Orig.            & Deconv.           &     \\
            & (K)            &  (AU)        & (AU)                &  (AU)                  &    \\
\hline
  $\alpha$ Men   &  $\leq$ 22     & $\geq$147    &92                   &  81                    &9.7$\times 10^{-7}$\\
  HD 88230   &  $\leq$ 22     & $\geq$62     &56                   & 51                     &1.6$\times 10^{-6}$\\   
  HD 210277  &         22     & 160          &$\leq$130            &                        &5.4$\times 10^{-6}$\\  
\hline

\end{tabular}
\end{center}
\end{table}

The  debris discs  in  this work  are a  new class  of discs
characterised by an  excess at 160 $\mu$m, little to no excess at
100 $\mu$m,  and no excess emission  at shorter wavelengths. These  discs are
the coldest  and least  luminous ones known  to date;
they  are significantly  colder and  fainter than  other  observed DUNES discs
(e.g.  Liseau et al.   2010, 
Marshall  et   al.  2011).  Their   
interpretation  poses significant challenges.

The shapes of the SEDs suggest that the dust is
located  in  a ring  with a larger inner void.
We have seen that the observed disc radii are smaller than 
those implied from $T_{\rm dust}$ (Table 4).
In other words,  $T_{\rm dust}$ is smaller than the black body temperature, 
$T_{\rm bb} \approx$ 25-30 K, given the observed sizes and  stellar 
luminosities.  Thus,   we  need  to  reconcile   these seemingly
contradictory results. Since one would reasonably expect a broad grain
size  distribution, which would  include small grains hotter  than
those indicated by the estimated  $T_{\rm dust}$, the question is how
to make the dust cold, while allowing a size distribution, but still keeping
the disc radii within the observed ones.

The low  temperatures of  grains require them  to be large  and highly
reflective. The required albedo can be derived from $T_{\rm dust}$ and
$T_{\rm bb}$.  Taking $T_{\rm dust}$  = 22 K as representative for the
three observed discs, and $T_{\rm bb}$  = 25-30 K, the albedo would be
$\gtrsim$50\%. 
 For comparison, surfaces of trans-Neptunian objects contain 
significant amounts of ice
(Barucci et al. 2011) and many, especially large ones (e.g.
Pluto, Eris), have albedos  in excess of 50\% (e.g. Vitense et al. 2010).
Accordingly,
Pluto has a surface brightness temperature  $\sim$ 10 K
below the black body value (Gurwell et al. 2010). It is natural
to expect that dust released from the surfaces of such objects would
have similar properties.

In addition, one has to explain why small grains in the cold discs are depleted.
One possibility is to assume a very low dynamical excitation of dust-producing 
planetesimals, so that discs would be devoid of small particles (Th\'ebault  \& Wu, 2008).
The reason is that low collision velocities between large grains, unaffected by
radiation pressure, create an imbalance between the rates at which small
grains are produced (low) and destroyed (high). Low collision velocities
are compatible with low orbital velocities at the large radii of the cold discs.
Besides, since the surface density of the solids far from the central stars is also low,
planetesimal accretion scenarios predict very long accretion timescales 
(Kenyon \& Bromley, 2008).  Thus, large planetesimals
that would  excite the discs may have failed to grow.
In fact, low dynamical excitation has been inferred for other
large debris discs, e.g.  HD 207129 (L\"ohne et al 2011).

The tenet that the cold discs are probably in a low dynamical excitation
state would  be difficult to reconcile with the existence
of large planets in the discs, since they would stir the discs too strongly.
The emission around $\alpha$~Men and HD~88230 is asymmetric, which might
be suggesting the presence of a giant planet as in Fomalhaut (Kalas et al.2008).
The problem can be mitigated if the planets are in nearly-circular
orbits or the planetesimals have low  eccentricities, as
suggested for  Fomalhaut (Chiang et al. 2009).
HD 210277 hosts a planet at 1.1~AU with $e_p= 0.47$;
however, based on the formulae by Mustill \& Wyatt (2009),
a stirring front from such a planet should not be able to reach the
$\sim 130$~AU-sized disc on  Gyr timescales.

Another point is the origin of the large inner voids in the cold discs.
These could either be due to clearing by planets
or may reflect the accretional and collisional history of primordial discs.
In the latter case, the observed radii  are those at which solids could reach
``right'' sizes, ``right'' degree of dynamical excitation, and/or were
able to survive over the stellar age, to produce the observed emission.
A detailed analysis of the nature of the cold discs is beyond the scope of
this paper, and we defer to Krivov et al. ({\it in prep.}) where possible scenarios
will be discussed in detail.

\section{Conclusions}

We have presented  $Herschel$ PACS observations of three  stars of the
 OTKP DUNES  sample. The observations reveal a  new class of
debris  discs with  fractional  luminosities
close to  the Solar-System Kuiper's  belt, but are colder  and larger.
These discs are a  challenge to current models explaining debris
discs around  mature solar-type stars such as  either the  usual  
collisional-dominated  disc  scenario or low dynamical excitation discs.


\end{document}